\documentclass[
aps,
twocolumn,
nofootinbib, 
prd, 
amsmath,amssymb,
notitlepage, 
longbibliography,
]{revtex4-2}

\usepackage[english]{babel}
\usepackage[utf8]{inputenc}
\usepackage[colorinlistoftodos]{todonotes}

\usepackage{hyperref}
\usepackage{url}
\usepackage{physics}

\hypersetup{
    colorlinks,
    linkcolor=blue,
    citecolor=blue,
    urlcolor=blue
}

\begin{document}

\title{Real scalar field, non-relativistic limit, and cosmological expansion}
\author{Lars H. Heyen}
\email{l.heyen@thhphys.uni-heidelberg.de}
\author{Stefan Floerchinger}
\email{stefan.floerchinger@thphys.uni-heidelberg.de}
    
\affiliation{Institut f\"{u}r theoretische Physik Heidelberg, Universit\"{a}t Heidelberg,\\ Philosophenweg 16, 69120 Heidelberg, Germany}

\begin{abstract}
The existing transformation from a relativistic real scalar field to a complex non-relativistic scalar field by Namjoo, Guth, and Kaiser is generalized from Minkowski space to a more general background metric. In that case the transformation is not purely algebraic any more but determined by a differential equation. We apply the generalized transformation to a real scalar with $\phi^4$ interaction on an Friedmann-Lema\^itre-Robertson-Walker cosmologically expanding background and calculate the resulting non-relativistic action up to second order in small parameters.
We also show that the transformation can be interpreted as a Bogoliubov transformation between relativistic and non-relativistic creation and annihilation operators and comment on emerging symmetries in the non-relativistic theory.
\end{abstract}
    
\maketitle
    
\section{Introduction}
The exact nature of dark matter is an open problem that reaches across multiple areas of physics from cosmology to particle physics. Compared to baryonic matter, the energy density of dark matter is about five times as large. Observational data implies that models of cold and collisionless dark matter are good candidates for a cosmological description \cite{PeeblesDarkMatter}. From the side of particle physics, a well-motivated candidate for dark matter particles are axions \cite{KJCG_Axions}.  Axions arise as the Goldstone bosons of a spontaneously broken $U(1)$ symmetry which is introduced to solve the strong CP problem and behave as pseudoscalars under Lorentz-transformations \cite{PQStrongCP, WeinbergAxion, WilczekAxion}. In the later universe we can expect such axions to have non-relativistic momenta \cite{Guth:2014hsa}.
Thus, a non-relativistic effective approach is appropriate. A non-relativistic limit of scalar field theories is needed in other contexts such as Bose-Einstein condensates and condensed-matter systems.

The precise relation between the relativistic theory and an effective non-relativistic description is important for interesting phenomenological questions, such as whether scalar field dark matter resembles a Bose-Einstein condensate or whether it could show superfluid behaviour \cite{Sikivie:2009qn, Saikawa:2012uk, Davidson:2013aba, Berges:2014xea, Davidson:2014hfa, Guth:2014hsa, Berezhiani:2015bqa, Braaten:2016kzc, Dev:2016hxv, Vikman:2017gxs, Rozner:2018ctk, Sharma:2018ydn, Braaten:2019knj, Ferreira:2020fam}.

Moreover, this question is important to understand the infrared regime in of relativistic quantum field theories and the corresponding universality classes. Specifically, relativistic scalar field theories can show condensation phenomena and one would like to understand in detail how they are related to various types of condensate in a non-relativistic field theoretic description \cite{Berges:2014bba, Moore:2015adu, Berges:2017ldx, Schmied:2018mte, Berges:2019oun}. 

Several different methods have been developed to arrive at an effective non-relativistic description for real scalars. 
In ref.\ \cite{RuffiniBonazolla}, Ruffini and Bonazzola developed a description for gravitationally bound, non-interacting bosons in Bose stars which was later generalized by Eby, Suranyi, and Wijewardhana in ref.\ \cite{Eby_2018}.
Mukaida, Takimoto and Yamada calculated an effective Lagrangian for the non-relativistic field by integrating out regions of the phase space not close to the mass pole in \cite{Mukaida2017}.
Their ansatz was shown in ref.\ \cite{BraatenMohapatraZhang} to be equivalent up to a field redefinition by Braaten, Mohapatra and Zhang to their effective Lagrangian from ref.\ \cite{Braaten:2016kzc}.
Namjoo, Guth, and Kaiser discovered an exact transformation between a relativistic real scalar field $\phi$ and a non-relativistic complex scalar field $\psi$ which has a Schroedinger-like equation of motion \cite{NamjooGuthKaiser}. This paper expands on their ideas by generalising their transformation to systems in curved space-time. For cosmological applications we are specifically interested in real scalars on a Friedmann-Lema\^{i}tre-Robertson-Walker (FLRW) metric. We managed to find a general class of transformations for a free real scalar which is minimally coupled to a classical background metric. The formalism preserves the invariance under coordinate transformations until the non-relativistic limit is taken and can naturally be extended to complex scalar fields. In FLRW space-time, introducing a potential term in the relativistic theory introduces a term proportional to $\Im(\psi^*\vec{\nabla}^2 \psi)$ in the effective potential for the non-relativistic field which not is not present on a time-independent background.

A kinetic description for real scalar fields in an expanding geometry has also been developed in refs.~\cite{Prokopec:2017ldn, Friedrich:2018qjv} while ref.\ \cite{Friedrich:2019zic} discusses a field theoretic approach and connects it to a kinetic description through the two-particle irreducible effective action formalism.

In section \ref{sec:PrevWork}, we give a quick overview over methods applied in order to obtain non-relativistic limits of theories in Minkowski space-time and point out problems that arise when applying this to real scalar fields on a non-Minkowskian background metric. In section \ref{sec:generalizedForm}, we introduce a generalization of the transformation proposed by Namjoo, Guth, and Kaiser in ref. \cite{NamjooGuthKaiser} to general Hamiltonian systems with linear equations of motion. Section \ref{sec:CovariantForm} covers a way to formulate this transformation in a covariant fashion which allows us to maintain invariance under coordinate transformations. In section \ref{sec:AppliedToFLRW}, this method is applied to a free real scalar on a Friedman-Lema\^{i}tre-Robertson-Walker background. Section \ref{sec:EffectivePot} contains the derivation of an effective non-relativistic potential for the same system with an added $\lambda\phi^4$ interaction in the relativistic theory. The natural extension of the formalism to complex scalar fields is discussed in section \ref{sec:ComplexScalars}. Section \ref{sec:Symmetries} deals with the symmetries that emerge in the transformed theory and their breaking under the addition of interaction terms. In section \ref{sec:Bogoliubov}, we interpret the non-relativistic limit as a Bogoliubov transformation between the annihilation and creation operators of relativistic and non-relativistic particles.
        
In this paper we choose units such that $\hbar = 1$ and $c = 1$ in all sections except for section \ref{sec:CToInfty} in which the factors of $c$ are explicitly written out and eventually taken to the limit $c\rightarrow\infty$.

Note added: While preparing the present manuscript, we became aware of a recent preprint by Salehian, Namjoo and Kaiser \cite{Salehian:2020bon} with a similar aim. More specific, the relation between a real relativistic scalar field and a complex, non-relativistic scalar field in the context of a  cosmologically expanding space-time is discussed there, as well. However, ref.\ \cite{Salehian:2020bon} differs from ours in several aspects. While the authors of ref.\ \cite{Salehian:2020bon} agree with us that the non-local field transformation proposed in ref.\ \cite{NamjooGuthKaiser} is not directly applicable for an expanding space-time, they do not generalize it in the way we will describe below, but instead work with an approximate local transformation with the problem that rapidly oscillating terms still appear in the equations of motion. A more detailed comparison between the approaches is left for future work.

\section{Previous Works \label{sec:PrevWork}}
\subsection{The $c\rightarrow\infty$ limit \label{sec:CToInfty}}

A simple way to take the non-relativistic limit in a classical field theory for a {\it complex} relativistic scalar field $\phi$ is to take the limit $c \rightarrow \infty$ in the equations of motion. One may start from the equations of motion for a free field in the form
\begin{align}
\frac{1}{c^2} \partial_t^2 \phi - \vec{\nabla}^2 \phi + m^2 c^2 \phi = 0.
\label{eq:freeComplexScalarEOM}
\end{align}
One splits off an oscillation with frequency $m c^2$ to define the non-relativistic field $\psi$ through the relation

\begin{align}
\phi = \frac{1}{\sqrt{2m}} e^{-i m c^2 t} \psi.
\label{eq:relativisticNonRelativisticRelationMinkowski}
\end{align}
Substituting this into the eq.\ \eqref{eq:freeComplexScalarEOM} leads to
\begin{align}
\frac{1}{2mc^2} \partial_t^2 \psi - i \partial_t \psi - \frac{1}{2m} \vec{\nabla}^2 \psi = 0.
\label{eq:secondOrderEqPsiNR}
\end{align}
In the limit $c \rightarrow \infty$ one can drop the first term and ends up with the Schr\"{o}dinger equation for a free particle,
\begin{align}
-i \partial_t \psi  - \frac{1}{2m} \vec{\nabla}^2 \psi=0.
\label{eq:firstOrderEqPsiNR}
\end{align}
Note that we now have a first order differential equation in time so that the Cauchy initial data consist only of the field itself. In contrast, for the Klein-Gordon equation one must also specify the time derivative $\partial_t \phi$ at some initial time or on an appropriate Cauchy surface. This shows that the Schroedinger equation propagates less information than the complex Klein-Gordon equation. In terms of the quantized theory, during the transition from eq.\ \eqref{eq:secondOrderEqPsiNR} to eq.\ \eqref{eq:firstOrderEqPsiNR}, we have dropped the information about possible anti-particle excitations.
            
Similarly, we can start with a free complex scalar field in a Friedmann-Lema\^itre-Robertson-Walker (FLRW) space-time described by a scale factor $a(t)$. The equation of motion is
\begin{equation}
\frac{1}{c^2} \partial_t^2 \phi + \frac{3 H}{c^2} \partial_t \phi - a^{-2}\vec{\nabla}^2 \phi + m^2 c^2 \phi = 0 
\label{eq:realscalareomFLRW}
\end{equation}
where $H = \dot{a}/a$ is the Hubble rate. Using the same relation \eqref{eq:relativisticNonRelativisticRelationMinkowski} we are first led to
\begin{equation}
\frac{1}{2mc^2} \partial_t^2 \psi - i \left( 1+i \frac{3 H}{2m c^2} \right) \partial_t \psi - i \frac{3H}{2} \psi - \frac{1}{2 a^2 m} \vec\nabla^2 \psi = 0. 
\end{equation}
The formal limit $c\to\infty$ yields then a Schr\"{o}dinger equation with additional Hubble damping or dilution term, 
\begin{equation}
-i \partial_t \psi - \frac{1}{2a^2m} \vec{\nabla}^2 \psi - \frac{3}{2} i H \psi =0.
\label{eq:firstOrderEqPsiNR_FLRW}
\end{equation}
While the method above works fine for some applications, 
we encounter problems if we try to use it for real scalar fields. A common ansatz to transform between the real scalar $\phi$ and the complex non-relativistic field $\psi$ is
\begin{equation}
\phi = \frac{1}{\sqrt{2m}} \Re(e^{-i m c^2 t} \psi)~.
\end{equation}
This leads to an equation of motion for $\psi$ which contains terms proportional to $\psi^*$. These terms contain fast oscillations proportional to $e^{\pm 2 i m c^2 t}$ and are usually neglected because at macroscopic time scales they are expected to average to zero. 
To avoid these terms altogether, at least in the non-interacting theory, we turn to a modified transformation proposed in ref. \cite{NamjooGuthKaiser}.

\subsection{Exact transformations}
\subsubsection{Minkowski space}

For a free, real scalar field in Minkowski space-time we have the Lagrangian density (we choose now units with $c=1$)
\begin{equation}
\mathcal{L} = \frac{1}{2} \eta^{\mu\nu} (\partial_\mu \phi)(\partial_\nu \phi) - \frac{1}{2} m^2 \phi^2,
\end{equation}
the Hamiltonian density
\begin{equation}
\mathcal{H} = \frac{1}{2} \pi^2 + \frac{1}{2} (\vec{\nabla} \phi)\cdot(\vec{\nabla} \phi) + \frac{1}{2} m^2 \phi^2, \label{eq:FlatSpaceHamiltonian}
\end{equation}
and equations of motion
\begin{equation}
\dot{\phi} = \pi, \quad\quad\quad \dot{\pi} = (\vec{\nabla}^2 - m^2)\phi.
\label{eq:FlatSpaceEOM}
\end{equation}

To bring this problem into a form where the non-relativistic limit is more apparent, ref.\ \cite{NamjooGuthKaiser} proposed the transformation to a new complex field $\psi$,
\begin{equation}
\begin{split}
\label{eq:NGKTransformationFlatSpace}
\psi & = \sqrt{\frac{m}{2}} e^{+i m t}  \left( \mathcal{P}^\frac{1}{2} \phi + \frac{i}{m} \mathcal{P}^{-\frac{1}{2}} \dot \phi \right), \\
\psi^* & = \sqrt{\frac{m}{2}} e^{-i m t} \left( \mathcal{P}^\frac{1}{2} \phi - \frac{i}{m} \mathcal{P}^{-\frac{1}{2}} \dot \phi \right).
\end{split}
\end{equation}
This uses the non-local derivative operator
\begin{equation}
\mathcal{P} = \sqrt{1 - \frac{\vec{\nabla}^2}{m^2}} = 1 - \frac{\vec{\nabla}^2}{2m^2} - \frac{(\vec{\nabla}^2)^2}{8 m^4}+\ldots,
\label{eq:DefOperatorP}
\end{equation}
where the right hand side is actually a low energy expansion. It can often be truncated after the second term, which yields then the standard non-relativistic limit. 

With the identification $\pi=\dot \phi$, eq.\ \eqref{eq:NGKTransformationFlatSpace} can be understood as a \textit{canonical} transformation (see also section \ref{sec:generalizedForm} and appendix \ref{app:CanonicalTrans}). The two real first order equations of motion \eqref{eq:FlatSpaceEOM} are now combined into a single first order complex equation
\begin{align}
-i\dot{\psi} + m(\mathcal{P} - 1)\psi =0.
\label{eq:psiEOMMinkowski}
\end{align}
Note that to leading order in the expansion on the right hand side of eq.\ \eqref{eq:DefOperatorP} one recovers Schr\"{o}dingers equation from \eqref{eq:psiEOMMinkowski}. Also, the equations of motion for $\psi$ do not have any fast oscillating terms like $e^{2i m t}\psi^*$ for this derivation based on the transformation \eqref{eq:NGKTransformationFlatSpace}. For completeness, we note also the inverse relation to \eqref{eq:NGKTransformationFlatSpace},
\begin{equation}
\begin{split}
\phi = & \frac{1}{\sqrt{2m}} \mathcal{P}^{-\frac{1}{2}} \left( e^{-imt} \psi + e^{+imt} \psi^* \right), \\
\dot \phi = &-i\sqrt{\frac{m}{2}} \mathcal{P}^{\frac{1}{2}} \left( e^{-imt} \psi - e^{+imt} \psi^* \right).
\end{split}
\label{eq:eNGKTransformationFlatSpaceInverse}
\end{equation}

Note that the transformation \eqref{eq:NGKTransformationFlatSpace} and its inverse \eqref{eq:eNGKTransformationFlatSpaceInverse} are, as a consequence of the definition \eqref{eq:DefOperatorP}, non-local in space but local with respect to time. Mathematically, the operator $\mathcal{P}$ it is defined through an eigenvalue decomposition of the Laplace-Beltrami operator $\vec{\nabla}^2$. Under many circumstances, the latter has real and negative (or vanishing) eigenvalues.
        		
We will refer to \eqref{eq:NGKTransformationFlatSpace} as the Namjoo-Guth-Kaiser (NGK) transformation.

\subsubsection{Expanding space-time}
For some applications such as the description of axions as a dark matter candidate it is interesting to consider an expanding Friedmann-Lema\^{i}tre-Robertson-Walker (FLRW) geometry. 
The equation of motion for a real scalar field becomes eq.\ \eqref{eq:realscalareomFLRW}.
One may attempt to use a similar transformation as previously in eq.\ \eqref{eq:NGKTransformationFlatSpace} with the operator in \eqref{eq:DefOperatorP} replaced by 
\begin{equation}
\mathcal{P}_a = \sqrt{1 - \frac{\vec{\nabla}^2}{a^2 m^2}}.
\label{eq:DefOperatorPModified}
\end{equation}
However, this does not lead to an equation of motion involving $\phi$ only, but to one which still contains fastly oscillating terms. Essentially the reason is that \eqref{eq:DefOperatorPModified} has an explicit time dependence because $\dot a \neq 0$. This motivates the more general approach to the problem to which we turn next.

\section{More general ansatz}
\subsection{Time dependent formalism \label{sec:generalizedForm}}

Our goal in this section is to find a \textit{linear transformation} from the original real scalar field and its time derivative or conjugate momentum to a complex scalar field and its complex conjugate which allow for an easy interpretation of the non-relativistic limit. This means that the transformed field $\psi$ should have a \textit{Schr\"{o}dinger-like equation of motion}
\begin{align}
i\dot\psi = \hat O \psi,
\end{align}
where $\hat O$ is a linear operator. From the NGK transformation we can expect this transformation to be non-local in space. However, we want it to be \textit{local in time} even in more complex geometries than just Minkowski space-time. Furthermore, the transformation between the original scalar field and its conjugate momentum and the real and imaginary part of the transformed field should \textit{preserve the Poisson-brackets (or commutators in a quantized theory)}.
    	    
As a first step, we restrict ourselves to systems with Hamiltonians which lead to linear equations of motion and do not depend on the direction of spatial momenta. Without loss of generality we can assume that there are no mixed terms between fields and their conjugate momenta as we could otherwise get rid of such terms via a canonical transformation. The most general Hamiltonian under these conditions then has the form
\begin{equation}
H = \int \mathrm{d}^3 x \left\{ \frac{1}{2} \pi A(t, - \vec \nabla^2) \pi + \frac{1}{2} \phi  B(t, - \vec \nabla^2) \phi \right\},
\label{eq:HamiltonianGeneralAB}
\end{equation}
where $\phi$ and $\pi$ are the scalar field and its conjugate momentum and $A(t, \vec p^2)$ and $B(t, \vec p^2)$ are two operators  which are analytical expressions of the operator $\vec{p}^2 =  -\vec{\nabla}^2$. (Note that $A$ and $B$ commute.) The equations of motion are now
\begin{equation}
\dot{\phi} = A(t, - \vec \nabla^2) \pi , \quad\quad\quad \dot{\pi} = -B(t, - \vec \nabla^2) \phi.
\label{eq:ansatzPhiPi}
\end{equation}
As an example, for an expanding FLRW geometry one would have
$A=a^{-3}$ and $B=a^{3}(m^2-\vec \nabla^2/ a^2)$.	
    		
Let us now make the ansatz for a new, complex field $\psi$ as a linear combination of the real field $\phi$ and its conjugate momentum,
\begin{equation}
\psi = \alpha(t, - \vec \nabla^2) \phi + i \beta(t, - \vec \nabla^2) \pi,
\label{eq:defNewComplexField}
\end{equation}
where $\alpha$ and $\beta$ are also functions of time and of $\vec{p}^2$ in momentum space. We further assume that they are invertible. Because all appearing operators commute we can be ignorant about their ordering and express them all as functions of the eigenvalues $\vec{p}^2$ of $-\vec{\nabla}^2$. 
The time derivative of \eqref{eq:defNewComplexField} follows with \eqref{eq:ansatzPhiPi} as
\begin{equation}
\dot{\psi} = \left( \dot{\alpha}\alpha^{-1} - i\beta B \alpha^{-1} \right) \alpha\phi + \left( \dot{\beta}\beta^{-1} - i \alpha A \beta^{-1} \right) i\beta\pi.
\label{eq:eompsi}
\end{equation}
We demand that \eqref{eq:eompsi} can be written as $\dot{\psi}\sim \psi$ to avoid fast-oscillating terms that appear together with $\psi^*$. It is convenient to define 
\begin{equation}
\gamma = \alpha^{-1}\beta,
\end{equation}
which is also a function of $\vec p^2$ and time $t$. From this definition and the Hamiltonian equations of motion, the proportionality we ask for becomes equivalent to the differential equation for time- and $\vec p^2$-dependent functions
\begin{equation}
\dot{\gamma}(t, \vec p^2) + i B(t, \vec p^2) \, \gamma(t, \vec p^2)^2 - iA(t, \vec p^2) = 0. 
\label{eq:gammaDEQ}
\end{equation}
Note that this equation is fully fixed by the Hamiltonian \eqref{eq:HamiltonianGeneralAB} although the solution needs in addition also an initial condition $\gamma(t_0, \vec p^2)$.
To determine the specific solution which allows for an easy interpretation as non-relativistic limit, we demand that the resulting transformation becomes the NGK transformation in the limit of Minkowski space.

For a $\gamma$ that solves eq.\ \eqref{eq:gammaDEQ} and $\psi = \alpha(\phi + i\gamma \pi)$, the equation of motion is then
\begin{equation}
\dot{\psi} = \left( \alpha^{-1}\dot{\alpha} - i B\gamma \right) \psi.     
\end{equation}
In the Minkowski space case we can read off from \eqref{eq:FlatSpaceHamiltonian} that $A=1$ and $B=(m^2 - \vec{\nabla}^2)$. This is indeed solved by $\gamma = m^{-1} \mathcal{P}^{-1} = B^{-1/2}$ so that $\dot{\gamma} = 0$, as was used in ref.~\cite{NamjooGuthKaiser}.
    		
The second important property that the transformation is supposed to have is that it should preserve the Poisson brackets (or equal time commutation relations).
To discuss this, we split $\psi$ into its real and imaginary parts with standard normalization as
\begin{equation}
\psi = \frac{1}{\sqrt{2}}(\varphi_1 + i \varphi_2).
\label{eq:psivarphi1varphi2}
\end{equation}
We want $\varphi_1$ and its conjugate momentum to have the same Poisson brackets as the original real scalar and its conjugate momentum. For a complex non-relativistic field with first order time derivative in the Lagrangian one expects that the conjugate momentum of the real part is proportional to the imaginary part and {\it vice versa}. With our normalization the conjugate momentum of $\varphi_1$ is $\varphi_2$ and the one of $\varphi_2$ is $-\varphi_1$.
    		
One way to ensure that the Poisson brackets are preserved is to demand that the transformation be \textit{canonical}. Here this means that it can be derived from a generating function $F_2$ via 
\begin{equation}
\pi(\phi, \varphi_2) = \frac{\partial F_2(\phi, \varphi_2)}{\partial \phi}, \quad
\varphi_1(\phi, \varphi_2) = \frac{\partial F_2(\phi, \varphi_2)}{\partial \varphi_2}.
\label{eq:generatingfunctionproperties}
\end{equation}
The Hamiltonian for the transformed field is then
\begin{equation}
\mathcal{H}_{\text{new}}(\varphi_1, \varphi_2) = \mathcal{H}_{\text{old}}(\phi(\varphi_1, \varphi_2), \pi(\varphi_1, \varphi_2)) + \frac{\partial F_2}{\partial t}.
\label{eq:transformationHamiltonian}
\end{equation}
    		
We now start with the transformation we already have and search for a generating function that matches it in order to confirm that the transformation is canonical. From equation \eqref{eq:defNewComplexField} and \eqref{eq:psivarphi1varphi2} we find
\begin{equation}
\begin{pmatrix} \phi \\ \pi \\ \end{pmatrix} = \frac{1}{\sqrt{2} |\alpha|^2 \Re(\gamma)}\begin{pmatrix} \Re(\alpha\gamma) & \Im(\alpha\gamma) \\ -\Im(\alpha) & \Re(\alpha) \\ \end{pmatrix}			
\begin{pmatrix}\varphi_1 \\ \varphi_2 \\ \end{pmatrix}.
\label{eq:phipiintermsofvarphi1varphi2}
\end{equation}
In order to find a generating function $F_2(\phi, \varphi_2)$ that fulfills eq.\ \eqref{eq:generatingfunctionproperties}
we need to first invert $\phi(\varphi_1, \varphi_2)$ for $\varphi_1$.   		
Rearranging \eqref{eq:phipiintermsofvarphi1varphi2} we get
\begin{equation}
\varphi_1 = \sqrt{2} \Re(\alpha\gamma)^{-1} \left(-\tfrac{1}{\sqrt{2}} \Im(\alpha\gamma) \varphi_2 + |\alpha|^2 \Re(\gamma)\phi \right). \label{eq:psiCtoOthers}
\end{equation}
Second, we need to express $\pi$ in terms of $\phi$ and $\varphi_2$,
\begin{equation}
\pi = \Re(\alpha\gamma)^{-1} \left(\tfrac{1}{\sqrt{2}} \varphi_2 - \Im(\alpha)\phi \right). \label{eq:pitoOthers}
\end{equation}
Integration of these expressions to get $F_2$ yields the condition for the transformation to be canonical, 
\begin{equation}
|\alpha|^2 = (2 \Re(\gamma))^{-1}. 
\label{eq:alphaCondition}
\end{equation}
In other words, once eq.\ \eqref{eq:gammaDEQ} has been solved, this fixes immediately also $|\alpha|$. The complex phase of $\alpha$ is still left undetermined, however.
As we will see later, this phase is related to a constant offset in the potential energy. 
Our choice in this paper will be $\arg(\alpha) = (m - V_0) t$.
A shift of the chemical potential is then realized as a phase shift of the non-relativistic field $\psi$ as (expected).
The transformation then takes the form
\begin{equation}
\begin{split}
\psi &= e^{i(m - V_0) t}(\Re(\gamma))^{-1/2} (\phi + i \gamma \pi)~,\\   
\psi^* &= e^{-i(m - V_0) t}(\Re(\gamma))^{-1/2} (\phi - i \gamma^* \pi)~.
\end{split}\label{eq:onlyGammaTrafo}
\end{equation}

The Poisson brackets / commutation relations 
are now related through
\begin{equation}
[\phi, \pi] = [\varphi_1, \varphi_2] = i [\psi, \psi^*].
\end{equation}

Now let us look at the influence of this condition on the equation of motion $\dot{\psi} = (\dot{\alpha}/\alpha- i B\gamma)\psi$. We can decompose $\alpha = |\alpha| e^{i (m - V_0) t}$, then
\begin{equation}
\frac{\dot{\alpha}}{\alpha} = \frac{\partial_t |\alpha|}{|\alpha|} + i (m - V_0),
\end{equation}
such that
\begin{equation}
\Re\left(\frac{\dot{\alpha}}{\alpha}\right) = \frac{\partial_t |\alpha|}{|\alpha|} = -\frac{1}{2}\frac{\partial_t \Re(\gamma)}{\Re(\gamma)}.
\end{equation}
    		
If we now assume that the eigenvalues of $A(t)$ and $B(t)$ are real, the real part of \eqref{eq:gammaDEQ} gives
\begin{equation}
B \Im(\gamma) = \frac{1}{2}\frac{\partial_t \Re(\gamma)}{\Re(\gamma)}.
\end{equation}
Thus, the real parts of $-iB\gamma$ and $\dot{\alpha}/\alpha$ exactly cancel which leaves us with the linear equation of motion 
\begin{equation}
i \dot{\psi} =  \left[B \Re(\gamma)- (m - V_0)  \right] \psi.
\label{eq:EOMpsi}
\end{equation}
This result is also in agreement with eq.\ \eqref{eq:transformationHamiltonian} (see also appendix \ref{app:CanonicalTrans}).
    		
The action corresponding to \eqref{eq:EOMpsi} is
\begin{equation}
S = \int_{t, \vec x}
\left\{ \frac{i}{2}(\dot{\psi}\psi^* - \psi \dot{\psi}^*) - \psi^* \left[ B\Re(\gamma) - (m - V_0)  \right] \psi \right\}.
\label{eq:Lagrangianpsi}
\end{equation}
Note that this is still an exact rewriting and equivalent to the Hamiltonian \eqref{eq:HamiltonianGeneralAB}. Given $A$ and $B$ one needs to solve eq.\ \eqref{eq:gammaDEQ} to determine $\gamma$ which in turn gives the transformation \eqref{eq:onlyGammaTrafo}. 
The real non-relativistic limit involves in addition also an expansion in orders of the Laplace-Beltrami operator.

As mentioned, for the Minkowski case we have $B = (m\mathcal{P})^2$ and $\gamma = \Re(\gamma) = (m\mathcal{P})^{-1}$ with $\mathcal{P}$ as defined in \eqref{eq:DefOperatorP}.
The lowest non-trivial order of the expansion in $\vec{\nabla}^2 / m^2$ is the Schroedinger equation with Hamiltonian $H = -\vec{\nabla}^2/(2m) + V_0$, higher orders add relativistic corrections.
Here we see that $V_0$ is indeed only a constant offset in the energy which can be changed at will, $V_0 \to V_0 + \Delta V_0$, by shifting the phase of the non-relativistic field, $\psi\to e^{-i\Delta V_0 t}\psi$.

\subsection{Covariant Formalism \label{sec:CovariantForm}}
While the above transformation works well in Minkowski space-time or an expanding cosmological space, time is somewhat singled out in the formalism and we would like to generalize this somewhat. We start from the action of a free massive real scalar field in a general space-time (using a mostly plus convention for the metric),
\begin{equation}
S = \int \mathrm{d}^4x \sqrt{-g} \left\{ -\frac{1}{2} g^{\mu \nu} (\nabla_\mu \phi) (\nabla_\nu \phi) - \frac{1}{2} m^2 \phi^2 \right\}.
\end{equation}
Then the Euler-Lagrange equation for $\phi$ is
\begin{equation}
-g^{\mu\nu}  \nabla_\mu \partial_\nu \phi + m^2 \phi = 0.
\label{eq:eomrealrelscalar}
\end{equation}
We now use a real, time-like vector field $u^\mu$ (normalized to $g_{\mu\nu} u^\mu u^\nu=-1$) to define $\psi$ as a linear combination of $\phi$ and its derivative $u^\mu\partial_\mu \phi$, 
\begin{equation}
\psi = \alpha \, \phi + i\beta \, u^\mu \partial_\mu \phi.
\end{equation}
We further define the projector perpendicular to $u^\mu$ as
\begin{equation}
\Delta_{\mu\nu} = g_{\mu\nu} + u_\mu u_\nu.
\end{equation}
The vector ``frame'' field $u^\mu$ can be used to define those velocities as non-relativistic that are mostly parallel to it, i.\ e.\ the velocity $v^\mu$ is called non-relativistic if
\begin{equation}
\Delta_{\mu\nu}v^\mu v^\nu / (u \cdot v)^2 \ll 1.
\end{equation}
Note that the frame field $u^\mu$ resembles somewhat a (relativistic) fluid velocity.

From the equation of motion \eqref{eq:eomrealrelscalar} we get
\begin{equation}
\begin{split}
u^\mu \partial_\mu \psi = & [u^\mu \partial_\mu \alpha + i\beta (-m^2 + \nabla_\mu \Delta^{\mu\nu} \partial_\nu)] \phi \\
& + [\alpha + i u^\nu \partial_\nu \beta - i \beta \nabla_\mu u^\mu] (u^\mu \partial_\mu \phi).
\end{split}
\end{equation}
Demanding that this be proportional to $\psi$ and introducing $\gamma = \alpha^{-1}\beta$, we end up with a differential equation for $\gamma$ which is very similar to \eqref{eq:gammaDEQ},
\begin{equation}
u^\nu \partial_\nu \gamma + i B \gamma^2 - (\nabla_\mu u^\mu) \gamma - i = 0,
\label{eq:gammaDEQNonFlat}
\end{equation}
where we have here
\begin{equation}
B = m^2 - \nabla_\mu \Delta^{\mu\nu} \partial_\nu.
\end{equation}
If we want $(\psi, \psi^*)$ to have the same commutation relations as $(\phi, u^\mu \partial_\mu \phi)$ up to a factor $i$, we end up with the same condition for $\alpha$, eq.\ \eqref{eq:alphaCondition}, and we can infer
\begin{equation}
B \Im(\gamma) = \frac{1}{2}\frac{u^\mu \partial_\mu \Re(\gamma)}{\Re(\gamma)} - \frac{1}{2} \nabla_\mu u^\mu.
\end{equation}

The resulting equation of motion for the complex field is now
\begin{equation}
 i \left[u^\mu \partial_\mu \psi + \frac{1}{2} (\nabla_\mu u^\mu) \psi \right] = - \left[u^\mu \partial_\mu \arg(\alpha) - B \Re(\gamma)\right] \psi.
\end{equation}
With the choice $\arg(\alpha) = (m - V_0)t$ this equation of motion corresponds to a Lagrangian very similar to the usual one for non-relativistic scalars,
\begin{equation}
\begin{split}
\mathcal{L} = &\frac{i}{2}\left[(u^\mu \partial_\mu \psi)\psi^* - \psi (u^\mu \partial_\mu \psi^*)\right] \\
&+ \psi^* \left[ u^\mu \partial_\mu \arg(\alpha) - B\Re(\gamma) \right] \psi.
\end{split}
\label{eq:LagrangianCovpsi}
\end{equation}
Going back to Minkowski space-time $g_{\mu\nu} = \eta_{\mu\nu}$, we can choose $u^\mu = (1, 0, 0, 0)$ and $\arg(\alpha) = (m - V_0) t$ which leads to the same differential equation for $\gamma$ as in the previous section and the action reduces to \eqref{eq:Lagrangianpsi}.

\subsection{Application to the FLRW case \label{sec:AppliedToFLRW}}
Now we can go back to the problem of an explicitly time-dependent space-time geometry. In FLRW space-time the metric is $g_{\mu\nu} = \text{diag}(-1, a^2, a^2, a^2)$
We now choose our frame fields as $u^\mu=(1,0,0,0)$ which fulfils the conditions from the previous section.
The the differential operator $u^\mu \nabla_\mu$ reduces to the time derivative in this frame (when acting on a scalar),
\begin{equation}
u^\mu \partial_\mu \psi = \partial_t \psi,
\end{equation}
and similarly
\begin{equation}
\nabla_\mu \Delta^{\mu\nu} \partial_\nu \psi = a^{-2}\vec{\nabla}^2 \psi.
\end{equation}
The next step is to solve differential equation \eqref{eq:gammaDEQNonFlat} using that $\nabla_\mu u^\mu = 3H$. This can be done by making the assumption that $\gamma$ can be written as a power series in $H$,
\begin{equation}
\gamma = \sum_{n=0}^\infty f_n(\mathcal{P}_a) H^n \label{eq:gammaAnsatz1},
\end{equation}
where the coefficients $f_n$ are analytical functions of the operator in \eqref{eq:DefOperatorPModified}.
The form of the $f_n$ is determined by a recursive formula discussed in appendix \ref{app:gammaExact}. There is an interplay of two frequency scales, one is given by the rest energy $m$ the other by the expansion of the universe or Hubble rate $\dot{a}/a = H$. While not strictly necessary, for simplicity we are going to make the assumption $H/m \ll 1$  corresponding to late times. The coefficient functions up to linear order in $H$ are
\begin{equation}
\begin{split}
&f_0(\mathcal{P}_a) = m^{-1} \mathcal{P}_a^{-1}, \\
&f_1(\mathcal{P}_a) = - \frac{iH}{2m^2} (\mathcal{P}_a^{-4} + 2 \mathcal{P}_a^{-2}).
\end{split}
\end{equation}
For the choice $V_0 = 0$
this gives us the transformation
\begin{equation}
\alpha = \sqrt{\frac{m}{2}} \mathcal{P}_a^{1/2} e^{i m t}, \quad \gamma = \frac{1}{m} \left[\mathcal{P}_a^{-1} -  \frac{iH}{2m} (\mathcal{P}_a^{-4} + 2\mathcal{P}_a^{-2})\right],
\end{equation}
such that 
\begin{equation}
\begin{split}
\psi = & \sqrt{\frac{m}{2}} \mathcal{P}_a^{1/2} e^{i m t}  \phi  \\
& +i  \frac{1}{\sqrt{2m}} \mathcal{P}_a^{1/2} e^{i m t}  \left[ \mathcal{P}_a^{-1} -  \frac{iH}{2m} \left(\mathcal{P}_a^{-4} + 2\mathcal{P}_a^{-2} \right)\right] \pi.
\end{split}
\end{equation}
The particular choice $V_0 = 0$ 
then leads to the equation of motion
\begin{equation}
i \left[ \dot{\psi} + \frac{3}{2} H \psi \right] = m (\mathcal{P}_a - 1)\psi.
\end{equation}
This is a Schr\"{o}dinger equation with a dilution term due to the expansion, sometimes called Hubble damping.

 \subsection{Effective Lagrangian for a $\lambda \phi^4$ theory \label{sec:EffectivePot}}
The following derivation of an effective Lagrangian and equation of motion for $\psi$ in an interacting theory follows closely the procedure  used by Namjoo, Guth, and Kaiser \cite{NamjooGuthKaiser}. 
For the relativistic theory we take the action as 
\begin{equation}
\begin{split}
S = \int \mathrm{d}^4x \, a^3 \Big{\{}&\frac{1}{2} (\partial_t \phi)^2 -\frac{1}{2} a^{-3} (\vec{\nabla} \phi)\cdot(\vec{\nabla} \phi) \\
&- \frac{1}{2} m^2 \phi^2 - \frac{\lambda}{4!} \phi^4 \Big{\}}.
\end{split}
\end{equation}
		        
The equation of motion for $\psi$ is
\begin{equation}
\begin{split}
i \left[\dot{\psi} + \frac{3}{2} H \psi \right] &= m (\mathcal{P}_a - 1)\psi + \frac{\lambda}{3!} \alpha \gamma ( \alpha^* \gamma^* \psi + \alpha \gamma \psi^*)^3 \\
&= m (\mathcal{P}_a - 1)\psi + \frac{\lambda}{4! m^2} \tilde{G},
\end{split}
\end{equation}
where we define
\begin{equation}
\begin{split}
\tilde{G} = & e^{i m t} \left(\mathcal{P}_a^{-1/2} - \frac{i H}{2 m}(\mathcal{P}_a^{-7/2} + 2 \mathcal{P}_a^{-3/2})\right) \\
& \times \left(\Psi e^{-i m t} + \Psi^* e^{i m t} \right)^3,
\end{split}
\end{equation}
and use
\begin{equation}
\Psi = \left(\mathcal{P}_a^{-1/2} + \frac{i H}{2 m}(\mathcal{P}_a^{-7/2} + 2 \mathcal{P}_a^{-3/2})\right) \psi.
\label{eq:PsipsiRelation}
\end{equation}
We split now the field $\psi$ into oscillations with different multiples of $\arg(\alpha) = mt$,
\begin{equation}
\psi = \sum_{\nu = -\infty}^\infty e^{i\nu m t} \psi_\nu.
\end{equation}
We also work with $\Psi_\nu$ which is related to $\psi_\nu$ linearly completely analogous to \eqref{eq:PsipsiRelation}. 
We also expand $\tilde{G}$ into a similar Fourier series,
\begin{equation}
\tilde{G} = \sum_{\nu = -\infty}^\infty e^{i\nu m t} \tilde{G}_\nu,
\end{equation}
and find then for the components
\begin{equation}
\begin{split}
\tilde{G}_\nu = &\left(\mathcal{P}_a^{-1/2} - \frac{i H}{2 m}(\mathcal{P}_a^{-7/2} + 2 \mathcal{P}_a^{-3/2})\right) \\
& \times \sum_{\mu, \mu^\prime = -\infty}^{\infty} {\Big \{} \Psi_\mu \Psi_{\mu^\prime} \Psi_{2 + \nu - \mu - \mu^\prime} + \Psi_\mu^* \Psi_{\mu^\prime}^* \Psi_{4 - \nu - \mu - \mu^\prime}^* \\
& \quad\quad\quad\quad + 3 \Psi_\mu \Psi_{\mu^\prime} \Psi_{\mu + \mu^\prime - \nu}^* + 
3 \Psi_\mu^* \Psi_{\mu^\prime}^* \Psi_{\nu - 2 + \mu + \mu^\prime} {\Big \}}.
\end{split}
\end{equation}
This lets us write equations of motion for each individual Fourier component $\psi_\nu$
\begin{equation}
i\left[\dot{\psi}_\nu + \frac{3}{2} H \psi_\nu \right] = m (\mathcal{P}_a - 1 + \nu) \psi + \frac{\lambda}{4! m^2} \tilde{G}_\nu,
\end{equation}
or similarly
\begin{equation}
\begin{split}
& i \left[\dot{\Psi}_\nu - K \Psi + \frac{3}{2} H \Psi_\nu \right] = m (\mathcal{P}_a - 1 + \nu) \Psi \\
& + \frac{\lambda}{4! m^2} \left(\mathcal{P}_a^{-1/2} + \frac{i H}{2 m}(\mathcal{P}_a^{-7/2} + 2 \mathcal{P}_a^{-3/2})\right) \tilde{G}_\nu,
\end{split}
\end{equation}
where 
\begin{equation}
K = \partial_t \mathrm{ln}\left(\mathcal{P}_a^{-1/2} + \frac{i H}{2 m}(\mathcal{P}_a^{-7/2} + 2 \mathcal{P}_a^{-3/2})\right),
\end{equation}
is proportional to the Hubble rate $H$.

This can be rewritten to give an expression for $\Psi_\nu$,
\begin{equation}
\Psi_\nu = -i \Gamma_\nu \left( \dot{\Psi}_\nu - K \Psi_\nu + \frac{3}{2} H \Psi_\nu \right) + \lambda G_\nu,
\end{equation}
with the abbreviation
\begin{equation}
\Gamma_\nu = -\left( m (\mathcal{P}_a - 1 + \nu) \right)^{-1},
\end{equation}
and
\begin{equation}
G_\nu = \frac{1}{4! m^2} \Gamma_\nu \left(\mathcal{P}_a^{-1/2} + \frac{i H}{2 m}(\mathcal{P}_a^{-7/2} + 2 \mathcal{P}_a^{-3/2})\right) \tilde{G}_\nu.
\end{equation}

For the non-relativistic limit we want to end up with an effective equation of motion for the slow field $\psi_s = \psi_{\nu = 0}$ and assume that all other $\psi_\nu$ and their time derivatives (corrected to account for the spatial expansion) are small. For this purpose we expand $\Psi_\nu$ and $G_\nu$ in orders of several small parameters namely $\lambda$, ($\dot{\Psi}_\nu - K\Psi_\nu + (3/2)H\Psi_\nu) / \Psi_\nu$, and eventually also $\vec{\nabla}^2 / m^2$.

We write formally the Fourier components of the field as a perturbative series
\begin{equation}
\Psi_\nu = \sum_{n = 0}^\infty \Psi_\nu^{(n)},
\end{equation}
where at leading order only the slow field $\Psi_s$ is non-vanishing, $\Psi_\nu^{(0)} = \delta_{\nu, 0} \Psi_s$.
Moreover, the slow field has no contributions from higher orders in the perturbative expansion, $\Psi_0^{(n > 0)} = 0$. The first order in the perturbative expansion gives for $\nu\neq0$ the relation 
\begin{equation}
\Psi_\nu^{(1)} = \lambda G_\nu^{(0)}, \label{eq:PsiNuFirstOrder}
\end{equation}
Higher orders are then governed by a recursive formula,
\begin{equation}
\Psi_\nu^{(n > 1)} = -i \Gamma_\nu \left( \dot{\Psi}_\nu - K \Psi_\nu + \frac{3}{2} H \Psi_\nu \right)^{(n - 1)} + \lambda G_\nu^{(n - 1)}.
\end{equation}
While it might seem problematic that there is $\dot{\Psi}_\nu$ on the right hand side, a term like this also arises in the Minkowski space-time case and was discussed in \cite{NamjooGuthKaiser} (and shown to be unproblematic in the appendix).
We use \eqref{eq:PsiNuFirstOrder} to expand the equation of motion for $\psi_s$ up to second order in small parameters (except for the spatial derivatives). For simplicity, we consider only terms up to linear order in $(H/m)$. This yields
\begin{equation}
\begin{split}
& i\left[\dot{\psi}_s + \frac{3}{2}H \psi_s\right] = m (\mathcal{P}_a - 1) \psi_s \\
& \quad\quad+ \lambda \Gamma_0^{-1} \left(\mathcal{P}_a^{-1/2} + \frac{i H}{2 m}(\mathcal{P}_a^{-7/2} + 2 \mathcal{P}_a^{-3/2})\right)^{-1} G_0 \\
&\approx m (\mathcal{P}_a - 1) \psi_s \\
& \quad\quad + \frac{\lambda}{8 m^2} \left(\mathcal{P}_a^{-1/2} - \frac{i H}{2 m}(\mathcal{P}_a^{-7/2} + 2 \mathcal{P}_a^{-3/2})\right) |\Psi_s|^2 \Psi_s \\
&\quad\quad + \frac{\lambda^2}{192 m^4} (3 \Gamma_2 + 6 \Gamma_2^* + \Gamma_4^* + \Gamma_{-2}) |\psi_s|^4 \psi_s.
\end{split}
\end{equation}
Expanding also in the spatial derivatives 
then yields
\begin{equation}
\begin{split}
& i\left[\dot{\psi}_s + \frac{3}{2} H \psi_s\right] \approx - \frac{1}{2a^2 m}\vec{\nabla}^2\psi_s + \frac{\lambda}{8 m^2} |\psi_s|^2 \psi_s \\
& - \frac{1}{8 a^4 m^3} \nabla^4 \psi_s \\
&+ \frac{\lambda}{32 a^2 m^4} \left[ \psi_s^2 \vec{\nabla}^2\psi_s^* + 2 |\psi_s|^2 \vec{\nabla}^2 \psi_s + \vec{\nabla}^2 (|\psi_s|^2 \psi_s) \right] \\
&+ i\frac{7 \lambda H}{32 a^2 m^5} \left[ -\psi_s^2 \vec{\nabla}^2\psi_s^* + 2 |\psi_s|^2 \vec{\nabla}^2 \psi_s - \vec{\nabla}^2 (|\psi_s|^2 \psi_s) \right] \\
&- \frac{17 \lambda^2}{768 m^5} |\psi_s|^4 \psi_s.
\end{split}
\end{equation}
This equation of motion then corresponds to an effective Lagrangian
\begin{equation}
\begin{split}
\mathcal{L}_{\text{eff}} = & \frac{i}{2} (\dot{\psi}_s \psi_s^* - \psi_s \dot{\psi}_s^*) - \frac{1}{2a^2 m}(\vec{\nabla}\psi_s)(\vec{\nabla}\psi_s^*) \\
& - \frac{\lambda}{16 m^2} |\psi_s|^4 + \frac{1}{8 a^4 m^3} (\vec{\nabla}^2\psi_s)(\vec{\nabla}^2\psi_s^*) \\
&  - \frac{\lambda}{32 a^2 m^4} |\psi_s|^2 (\psi_s^* \vec{\nabla}^2 \psi_s + \psi_s \vec{\nabla}^2 \psi_s^*) \\
&- i\frac{7 \lambda H}{32 a^2 m^5} |\psi_s|^2 (\psi_s^* \vec{\nabla}^2\psi_s - \psi_s \vec{\nabla}^2\psi_s^*) \\
& + \frac{17 \lambda^2}{9\cdot 2^8 m^5} |\psi_s|^6.
\end{split}
\label{eq:effectiveLagrangianNR}
\end{equation}
Notably, we have a term with an imaginary coefficient in the effective Lagrangian. This term is proportional to both coupling strength $\lambda$ and the Hubble rate $H$. It does however not break unitarity because $(\psi_s^* \vec{\nabla}^2\psi_s - \psi_s \vec{\nabla}^2\psi_s^*) = 2i \Im(\psi_s^* \vec{\nabla}^2\psi_s)$ is purely imaginary.

Let us note that while the last term \eqref{eq:effectiveLagrangianNR} seems to imply an instability (because the coefficient of $|\psi_s|^6$ is positive and therefore contributes negatively to the effective potential), this is not necessarily a problem. An additional term arises from a corresponding term in the relativistic theory $\sim\phi^6$ that can counterbalance this term. 
Such stabilizing terms might well be contained in higher orders of the expansion of the cosine potential usually assumed for the axion.

\subsection{Complex scalar fields \label{sec:ComplexScalars}}
Let us now turn to complex relativistic fields and study their non-relativistic limit, as well.
We may decompose a complex scalar field in terms of two real fields,
\begin{equation}
\Phi = \frac{1}{\sqrt{2}}(\phi_1 + i \phi_2).
\end{equation}
The behaviour of the real scalar fields $\phi_1$ and $\phi_2$ is given by the Hamiltonian dynamics in eq.\ \eqref{eq:ansatzPhiPi}. For simplicity, let us discuss two non-interacting real scalars in Minkowski space-time such that the Lagrangian can be written in terms of the complex field,
\begin{equation}
\mathcal{L} = -\eta^{\mu\nu} (\partial_\mu \Phi^*) (\partial_\nu \Phi) - m^2 \Phi^* \Phi.
\end{equation}
The conjugate momenta for the complex scalar and its complex conjugate are
\begin{equation}
\Pi = \dfrac{\partial \mathcal{L}}{\partial \dot{\Phi}} = \dot{\Phi}^*, \quad\quad\quad
\Pi^* = \dfrac{\partial \mathcal{L}}{\partial \dot{\Phi}^*} = \dot{\Phi}.
\end{equation}
The Hamiltonian assumes the form
\begin{equation}
\mathcal{H} = \Pi^* \Pi + (\vec{\nabla} \phi^*)(\vec{\nabla} \phi) + m^2 \Phi^* \Phi.
\end{equation}
We now transform the real scalars $\phi_1$ and $\phi_2$ into (non-relativistic) complex fields $\psi_1$ and $\psi_2$ respectively, following the same steps as previously. The Lagrangian becomes 
\begin{equation}
\mathcal{L} = \sum_{n = 1}^{2}\frac{i}{2}(\dot{\psi}_n\psi_n^* - \psi_n \dot{\psi}_n^*) - \psi_n^* \left( m (\mathcal{P} - 1) + V_0 \right) \psi_n.
\end{equation}
Similarly to the transformation for the real scalars we define
\begin{equation}
\begin{split}
\Psi_1 &= \alpha (\Phi + i\gamma \Pi^*) = \frac{1}{\sqrt{2}}(\psi_1 + i \psi_2), \\
\Psi_2 &= \alpha (\Phi^* + i\gamma \Pi) = \frac{1}{\sqrt{2}}(\psi_1 - i \psi_2).
\end{split}
\end{equation}
These two fields are for particles and anti-particles, respectively. 
The Lagrangian can be rewritten in terms these fields as
\begin{equation}
\begin{split}
\mathcal{L} = & \sum_{n = 1}^{2} {\Big \{} \frac{i}{2}(\dot{\Psi}_n\Psi_n^* - \Psi_n \dot{\Psi}_n^*) \\
& \quad\quad- \Psi_n^* \left( m (\mathcal{P} - 1) + V_0 \right) \Psi_n {\Big \}}.
\end{split}
\end{equation}
While we used Minkowski space-time as an example, the transformation and resulting Lagrangian holds in general for systems in which the transformation of the real scalars can be performed.

In covariant notation,
\begin{equation}
\begin{split}
\Psi_1 &= \alpha (\Phi + i\gamma u^\mu \nabla_\mu \Phi) = \frac{1}{\sqrt{2}}(\psi_1 + i \psi_2), \\
\Psi_2 &= \alpha (\Phi^* + i\gamma u^\mu \nabla_\mu \Phi^*) = \frac{1}{\sqrt{2}}(\psi_1 - i \psi_2). \\
\end{split}
\label{eq:complexScalarTrafo}
\end{equation}
The Lagrangian becomes
\begin{equation}
\begin{split}
\mathcal{L} = & \sum_{n = 1}^{2} {\Big \{}\frac{i}{2}((u^\mu \partial_\mu \Psi_n) \Psi_n^* - \Psi_n (u^\mu \partial_\mu \Psi_n^*)) \\
& \quad\quad + \Psi_n^* \left( u^\mu \partial_\mu \arg(\alpha) - B\Re(\gamma) \right) \Psi_n {\Big\}}.
\end{split}
\end{equation}
While a non-interacting relativistic theory will always lead to a Lagrangian that decomposes into separate $\Psi_1$ and $\Psi_2$ parts for particles and anti-particles respectively, interactions in the relativistic theory will introduce mixed terms into the non-relativistic Lagrangian.
    		
To lowest order in $\vec{\nabla}^2$ the equation of motion matches with the simpler limit from the introduction for Minkowski and FLRW space-time if we make the choices $\arg(\alpha) = m t$, $u^\mu = (1, 0, 0, 0)$ and set $\Psi_2(x) = 0$.
This last choice is the point at which we decide to neglect any possible antiparticle excitations which are not present in the simpler limit.
For Minkowski space-time in lowest order approximation the equation of motion becomes the one obtained from the naive formalism \eqref{eq:firstOrderEqPsiNR},
    		
\subsection{Symmetries \label{sec:Symmetries}}
The Lagrangian of the free real scalar field has for a general space-time metric a $\mathbb{Z}_2$ symmetry. In the transformed Lagrangian \eqref{eq:Lagrangianpsi} this symmetry is upgraded to a global $U(1)$ symmetry. The associated Noether charge is
\begin{equation}
Q = \int \mathrm{d}^3 x \sqrt{-g} |\psi|^2, \quad\quad\quad \dot{Q} = 0.
\end{equation}
For Minkowski space-time we have something that resembles the conservation of the particle number in quantum mechanics.
In FLRW space-time the determinant of the metric introduces a factor that accounts for the spatial expansion over time which implies $\int \mathrm{d}^3 x |\psi|^2 \propto a^{-3}(t)$.
 
 Note that this emergent $U(1)$ symmetry is in principle broken by any interaction term $\phi^n$ in the original Lagrangian.
However, the low energy perturbative expansion we have made in section \ref{sec:EffectivePot} enforces it even in the approximation to the interacting theory. Beyond this approximation, inelastic processes such as $2\to 4$ and $4\to 2$ should be possible that also break the emergent $U(1)$ symmetry.

In the case of the complex scalar field, the Lagrangian has a global $U(1)$ symmetry already before the transformation.
After the transformation there are two independent $U(1)$ symmetries for $\Psi_1$ and $\Psi_2$ each as well as a $U(2)$ symmetry for the duplet $(\Psi_1, \Psi_2)$.
The $U(1)$ symmetries again correspond to something akin to conserved individual particle numbers.
If there is an interaction term present in the original action that preserves the $U(1)$ symmetry of $\Phi$, the individual $U(1)$ symmetries for $\Psi_1$ and $\Psi_2$ as well as the $U(2)$ symmetry break down to a single global $U(1)$ symmetry realized by
\begin{equation}
\Psi_1 \to e^{i\alpha} \Psi_1, \quad\quad\quad  \Psi_2 \to e^{-i\alpha} \Psi_2.
\end{equation}
The corresponding Noether charge then is
\begin{equation}
 Q = \int \mathrm{d}^3 x (|\Psi_1|^2 - |\Psi_2|^2), \quad\quad\quad \dot{Q}= 0,
\end{equation}
which resembles charge conservation. Notably, this symmetry does not have to be explicitly enforced in the approximation to be maintained in the non-relativistic limit.

\subsection{Transformation of the functional integral}
We start from the Hamiltonian form of the path integral for the relativistic scalar theory in Minkowski space,
\begin{equation}
 Z = \int D\phi \, D\pi \exp \left\{ i\int d^4x \, [\pi \dot{\phi} - \mathcal{H}(\phi, \pi)] \right\}.
\end{equation}
This form of the functional integral is most directly related to a non-relativistic functional integral
\begin{equation}
 Z = \int D\psi \, D\psi^* \exp \left\{ i S[\psi, \psi^*] \right\}.
\end{equation}
Indeed, the complex field $\psi$ is just a linear combination of $\phi$ and the conjugate momentum field $\pi$ as displayed by eq.\ \eqref{eq:defNewComplexField}, cf.\  also ref.\ \cite{Gollisch:2000jh}.
We use Fujikawa's method (cf.\ \cite{Bertlmann:AnomaliesInQFT}) to identify potential anomalies.
In order for the transformation to possibly be free of anomalies, we must show that the transformation of the integral measure does not add further terms.
For this purpose we decompose both relativistic and non-relativistic fields into orthonormal eigenfunctions $\hat{\lambda}_n$ of the Laplace-Beltrami operator $-\vec{\nabla}^2$ with eigenvalues $\lambda_n$ as
\begin{equation}
\begin{split}
 \phi &= \sum_n a_n(t) \hat{\lambda}_n,\quad\quad\quad \pi = \sum_n b_n(t) \hat{\lambda}_n,\\
 \psi &= \sum_n a^\prime_n(t) \hat{\lambda}_n,\quad\quad\quad \psi^* = \sum_n b^\prime_n(t) \hat{\lambda}_i.
\end{split}
\end{equation}
The transformation of the path integral measure is given by the Jacobian of the tranformation between primed and unprimed coefficients.
The connection between the relativistic and non-relativistic coefficients is given by 
\begin{equation}
\begin{split}
a^\prime_n(t) &= \hat{\lambda}_n^\dagger \cdot \psi = \alpha(\lambda_n)[a_n(t) + i\gamma(\lambda_n) b_n(t)],\\
b^\prime_n(t) &= \hat{\lambda}_n^\dagger \cdot \psi^* = \alpha^*(\lambda_n)[a_n(t) - i \gamma^*(\lambda_n) b_n(t)].
\end{split}
\end{equation}
The Jacobian of the transformation is now the determinant of a block matrix
\begin{equation}
\begin{split}
J &= \Bigg{|}\det \begin{pmatrix}
\frac{\partial a_m^\prime}{\partial a_n} & \frac{\partial a_m^\prime}{\partial b_n} \\
\frac{\partial b_m^\prime}{\partial a_n} & \frac{\partial b_m^\prime}{\partial b_n}
\end{pmatrix}\Bigg{|}\\
&= \Big{|}\det \begin{pmatrix}
\delta_{mn} \alpha(\lambda_n) & i\delta_{mn} \alpha(\lambda_n)\gamma(\lambda_n) \\
\delta_{mn} \alpha^*(\lambda_n) & -i\delta_{mn} \alpha^*(\lambda_n)\gamma^*(\lambda_n)
\end{pmatrix}\Big{|}\\
&= |\det (-i \delta_{mn} |\alpha(\lambda_n)|^2 (\gamma^*(\lambda_n) + \gamma(\lambda_n)))| \\
&= |\det (-i \delta_{mn})| = 1.
\end{split}
\end{equation}
We infer that the measure of the path integral seems to have no obvious anomalies under the NGK transformation, at least for the charge neutral fields we have investigated.

\section{Interpretation as Bogoliubov transformation \label{sec:Bogoliubov}}
In Minkowski space-time it is possible to decompose the scalar field operator into annihilation and creation operator as
\begin{equation}
\phi = \int \frac{\mathrm{d}^3 k}{(2\pi)^3 \sqrt{2E_{\vec{k}}}} (a_{\vec{k}} e^{i \vec{k} \vec{x}} + a_{\vec{k}}^\dagger e^{-i \vec{k} \vec{x}}).
\end{equation}
In a time-dependent situation such as a cosmological expansion one can write similarly
\begin{equation}
\phi = \int \frac{\mathrm{d}^3 k}{(2\pi)^3} (a_{\vec{k}} f_{\vec{k}}(t) e^{i \vec{k} \vec{x}} + a_{\vec{k}}^\dagger f_{\vec{k}}^*(t) e^{-i \vec{k} \vec{x}})~,
\end{equation}
but there is an additional freedom in the choice of the mode functions $f_{\vec{k}}$. The latter are only restricted by a differential equation and a normalization condition. Different choices of mode functions correspond to different choices of annihilation and creation operators and also have different vacua.

This change between operators can be expressed as a linear transformation
\begin{equation}
 b_{\vec{k}} = u_{\vec{k}} a_{\vec{k}} + v_{-\vec{k}}^* a_{-\vec{k}}^\dagger, \quad\quad\quad
 b_{\vec{k}}^\dagger = u_{\vec{k}}^* a_{\vec{k}}^\dagger + v_{-\vec{k}} a_{-\vec{k}}. \label{eq:BogoliubovTrf}
\end{equation}
	    If the new set of operators $(b_{\vec{k}}, b_{\vec{k}}^\dagger)$ can be written in this way in terms of the old ones $(a_{\vec{k}}, a_{\vec{k}}^\dagger)$ and the condition $|u_{\vec{k}}|^2 - |v_{-\vec{k}}|^2 = 1$ holds, this is called a Bogoliubov transformation. This type of transformation leaves the commutation relations invariant, $[b_{\vec{k}}, b_{\vec{k}}^\dagger] = [a_{\vec{k}}, a_{\vec{k}}^\dagger]$.

However, a vacuum state with respect to a certain choice of operators might not be vacuum state with respect to the transformed ones. The density of $b$-particles in the $a$-vacuum is given by $|v_{\vec{k}}|^2$.
As we are working in the context of quantum fields on curved backgrounds, it is interesting to express our transformation between relativistic and non-relativistic fields as a Bogoliubov transformation.
	    
\subsection{Minkowski space-time}
The canonical transformation introduced can also be interpreted as a Bogoliubov transformation between two sets of annihilation and creation operators $(a_{\vec{p}}, a_{\vec{p}}^\dagger)$ and $(b_{\vec{p}}, b_{\vec{p}}^\dagger)$. To see this we start with the standard definition of these operators for a real scalar in Minkowski space-time,
\begin{equation}
\begin{split}
\phi &= \int \frac{\mathrm{d}^3 p}{(2\pi)^3} \frac{1}{\sqrt{2 E_{\vec{p}}}} \left( a_{\vec{p}} e^{i \vec{p} \vec{x}} + a_{\vec{p}}^\dagger e^{-i \vec{p} \vec{x}} \right), \\
\pi &= - i \int \frac{\mathrm{d}^3 p}{(2\pi)^3} \sqrt{\frac{E_{\vec{p}}}{2}} \left( a_{\vec{p}} e^{i \vec{p} \vec{x}} - a_{\vec{p}}^\dagger e^{-i \vec{p} \vec{x}} \right),
\end{split}
\end{equation}
with $E_{\vec{p}} = \sqrt{\vec{p}^2 + m^2}$. We then define the creation and annihilation operators for the transformed fields through
\begin{equation}
\psi = \int \frac{\mathrm{d}^3 p}{(2\pi)^3} b_{\vec{p}} e^{i \vec{p} \vec{x}}, \quad\quad\quad
\psi^* = \int \frac{\mathrm{d}^3 p}{(2\pi)^3} b_{\vec{p}}^\dagger e^{-i \vec{p} \vec{x}}.
\end{equation}
Using the transformation $\psi = \alpha (\phi + i\gamma \pi)$ and comparing yields relation \eqref{eq:BogoliubovTrf} between operators
with the mode functions
\begin{equation}
\begin{split}
&u_{\vec{p}} = \frac{1}{\sqrt{2 E_{\vec{p}}}} (\alpha + i\alpha\gamma E_{\vec{p}}), \\
&v_{\vec{p}} = \frac{1}{\sqrt{2 E_{\vec{p}}}} (\alpha^* + i \alpha^* \gamma^* E_{\vec{p}}) = v_{-\vec{p}},
\end{split}
\end{equation}
where $\alpha$ and $\gamma$ are now no longer differential operators, but each $\vec{\nabla}^2$ is replaced by $-p^2$. 
To confirm that the transformation indeed leads to the correct commutation relations for the new operators we have to check
\begin{equation}
|u_{\vec{p}}|^2 - |v_{\vec{p}}|^2 = |\alpha|^2 (\gamma^* + \gamma) = 1,
\end{equation}
which is equivalent to condition \eqref{eq:alphaCondition}. A consequence of the Bogoliubov transformation is that a vacuum state defined w.r.t. $(a_{\vec{p}}, a_{\vec{p}}^\dagger)$ is not in general a vacuum state w.r.t. $(b_{\vec{p}}, b_{\vec{p}}^\dagger)$. The number density of $b$-particles in the $a$-vacuum is
\begin{equation}
\expval{n_{b}}{0_a} = \int \frac{\mathrm{d}^3 p}{(2\pi)^3} |v_{\vec{p}}|^2,
\end{equation}
with
\begin{equation}
\begin{split}
|v_{\vec{p}}|^2 &= \Bigg{|}\frac{1}{\sqrt{2 E_{\vec{p}}}} (\alpha^* + i\alpha^* \gamma^* E_{\vec{p}})\Bigg{|}^2 \\
&= -\frac{1}{2} |\alpha|^2 (\gamma + \gamma^*) + \frac{|\alpha|^2}{2 E_{\vec{p}}} (1 + |\gamma|^2 E^2_{\vec{p}}).
\end{split}
\end{equation}
After applying condition \eqref{eq:alphaCondition}, the first term becomes $-1/2$. The second term requires us to substitute the actual transformation we chose. In the case of the real scalar in Minkowski space-time that means
\begin{equation}
|\alpha|^2 = \frac{1}{2} E_{\vec{p}}, \quad\quad\quad |\gamma|^2 = 1/E^2_{\vec{p}}.
\end{equation}
With this we find $v_p = 0$ which implies also that relativistic and non-relativistic theory have the same vacuum state, $\ket{0_a} = \ket{0_b}$, This result is very intuitive since it confirms that, even on the level of a quantum theory, the free theory before and after the transformation describe the same particle excitations.
	        
\subsection{Cosmologically expanding space-time}
In FLRW space-time the expansion of the relativistic fields is usually done with a mode function $f_{\vec{k}}$. We assume $f_{\vec{k}} = f_{-\vec{k}}$ and write
\begin{equation}
\begin{split}
\phi &= \frac{1}{\sqrt{2}} \int \frac{\mathrm{d}^3 k}{(2\pi)^3} \left[ f_{\vec{k}}(t) a_{\vec{k}} e^{i\vec{k} \vec{x}} + f_{\vec{k}}^*(t) a_{\vec{k}}^\dagger e^{-i\vec{k} \vec{x}} \right], \\
\pi &= \dot{\phi} = \frac{1}{\sqrt{2}} \int \frac{\mathrm{d}^3 k}{(2\pi)^3} \left[ \dot{f}_{\vec{k}}(t) a_{\vec{k}} e^{i\vec{k} \vec{x}} + \dot{f}_{\vec{k}}^*(t) a_{\vec{k}}^\dagger e^{-i\vec{k} \vec{x}} \right].
\end{split}
\end{equation}
If we demand that both $(\phi, \pi = u^\mu \nabla_\mu \phi = \dot{\phi})$ and $(a_{\vec{k}}, a_{\vec{k}}^\dagger)$ fulfill the usual commutation relations
\begin{equation}
\begin{split}
[\phi(t, \vec{x}), \pi(t, \vec{y})] = i \delta(\vec{x} - \vec{y}), \quad\quad\quad
[a_{\vec{k}}, a_{\vec{p}}^\dagger] = \delta(\vec{k} - \vec{p}),
\end{split}
\end{equation}
we get the condition for the mode function
\begin{equation}
\Im(f_{\vec{k}}^*(t)~ \dot{f}_{\vec{k}}(t)) = -1.
\end{equation}
	            
Let now $U_{\vec{k}} \in GL(2, \mathbb{C})$ be an invertible linear transformation on which we will impose further conditions later and $\zeta_{\vec{k}}$ a function of time and momentum. We define $b_{\vec{k}}$ and $b_{\vec{k}}^\dagger$ through
\begin{equation}
\begin{pmatrix} \psi \\ \psi^* \\ \end{pmatrix} = \frac{1}{\sqrt{2}} \int \frac{\mathrm{d}^3 k}{(2 \pi)^3} U_{\vec{k}} \begin{pmatrix} \zeta_{\vec{k}}(t) f_{\vec{k}}(t) b_{\vec{k}} e^{i\vec{k} \vec{x}} \\ \zeta_{\vec{k}}^*(t) f_{\vec{k}}^*(t) b_{\vec{k}}^\dagger e^{-i\vec{k} \vec{x}} \\ \end{pmatrix}.
\end{equation}
Again, we demand that the standard equal time commutation relations hold,
\begin{equation}
[\psi(t, \vec{x}), \psi^*(t, \vec{y})] = \delta(\vec{x} - \vec{y}), \quad\quad\quad
[b_{\vec{k}}, b_{\vec{p}}^\dagger] = \delta(\vec{k} - \vec{p}).
\end{equation}
This imposes a constraint on the modulus of $\zeta_{\vec{k}}$,
\begin{equation}
 \frac{|f_{\vec{k}}|^2 \text{det}(U_{\vec{k}})}{2} |\zeta_{\vec{k}}|^2 = 1.
\end{equation}
After substituting $\psi = \alpha(\phi + i\gamma \pi)$ and inverting $u_{\vec{k}}$, we identify transformation coefficients between $(a_{\vec{k}}, a_{\vec{k}}^\dagger)$ and $(b_{\vec{k}}, b_{\vec{k}}^\dagger)$ as in the Minkowski space-time case \eqref{eq:BogoliubovTrf}.
In order for this to possibly be a Bogoliubov transformation, the matrix $U_{\vec{k}}$ has to obey
\begin{equation}
\begin{split}
&(U_{\vec{k}}^{-1})_{12} = (U_{\vec{k}}^{-1})_{21}^* =\nu^*, \\
&(U_{\vec{k}}^{-1})_{11} = (U_{\vec{k}}^{-1})_{22}^* = \mu.
\end{split}
\end{equation}
With this the transformation coefficients can be expressed as
\begin{equation}
\begin{split}
&u_{\vec{k}} = (\zeta_{\vec{k}} f_{\vec{k}})^{-1} [\mu \alpha (f_{\vec{k}} + i\gamma\dot{f}_{\vec{k}}) + \nu^* \alpha^* (f_{\vec{k}} - i\gamma^* \dot{f}_{\vec{k}})], \\
&v_{\vec{k}} = (\zeta_{\vec{k}}^* f_{\vec{k}}^*)^{-1} [\nu \alpha (f_{\vec{k}} + i\gamma\dot{f}_{\vec{k}}) + \mu^* \alpha^* (f_{\vec{k}} - i\gamma^* \dot{f}_{\vec{k}})].
\end{split}
\end{equation}
The last requirement for this to be a Bogoliubov transformation is $|u_{\vec{k}}|^2 - |v_{\vec{k}}|^2 = 1$. One may check that this is indeed fulfilled. 
For an appropriate choice of $U_{\vec{k}}$ and $\zeta_{\vec{k}}$ the transformation from $(a_{\vec{k}}, a_{\vec{k}}^\dagger)$ to $(b_{\vec{k}}, b_{\vec{k}}^\dagger)$ can be interpreted as a Bogoliubov transformation, indeed. Notably, the conditions imposed on $U_{\vec{k}}$ when combined with the normalization $\text{det}(U) = 1$ constrain $U_{\vec{k}}$ exactly to the group of Bogoliubov transformations.
	            
\section{Conclusions}
In this paper, we have extended the methods for taking the non-relativistic limit of a field theory developed by Namjoo, Guth, and Kaiser \cite{NamjooGuthKaiser} to more general space-time metric.
This requires introducing a normalized timelike vector that defines non-relativistic velocities as being parallel to it except for small deviations. One is then led to a differential equation, the solution of which fixes the precise form of the canonical transformation between (real) relativistic and (complex) non-relativistic fields.

The method we developed allows the Lagrangian of the non-relativistic field to still be invariant under general coordinate transformations at an intermediate step. This invariance is then lost later in the actual low energy approximation.
The formalism extends naturally to complex scalar fields and leads to a non-relativistic description of both particles and antiparticles. 
Neglecting either one and expanding only to lowest non-trivial order in momenta however agrees  with the more naive $c\rightarrow\infty$ limit;  this is a good check of consistency.
		
We applied the generalized transformation to a real scalar field in FLRW space-time.
This introduces a non-vanishing imaginary part in the prefactor $\gamma$ of the momentum field 
that is not present in Minkowski space-time.
The mechanism for obtaining an effective theory for the non-relativistic field still works very similar in an expanding FLRW space-time.
In this geometry we calculated the effective non-relativistic action corresponding to a $\phi^4$-interaction in the relativistic theory up to second order in small parameters.
The main differences compared to Minkowski space-time are 
the scale factors accompanying each spatial derivative and a term proportional to $\Im(\psi^*\vec{\nabla}^2 \psi)$ in the effective non-relativistic action which is caused by the imaginary part of the transformation parameter~$\gamma$.
		
For free theories the transformation from relativistic to non-relativistic description causes symmetries to emerge or to extend.
For real scalars, a global $U(1)$ symmetry of the non-relativistic field emerges while for complex scalars the $U(1)$ symmetry is extended to a $U(2)$ symmetry.
These emergent symmetries are in general broken by interaction terms.
The most direct low energy approximation scheme for obtaining the effective non-relativistic action however preserves the $U(1)$ symmetry of the non-relativistic field corresponding to a relativistic real scalar.
The associated conserved charge is the spatial integral over the particle density $\rho =|\psi_s|^2$ in Minkowski space-time. 
The density of non-relativistic particles in expanding space-time behaves as expected, scaling as $\rho\sim a^{-3}$.
The $U(1)$ symmetry of a complex relativistic scalar translates into a conservation of the difference of particle and antiparticles numbers $|\Psi_1|^2 - |\Psi_2|^2$ which resembles charge conservation, independent of low energy approximations.
		
Furthermore, we showed that, provided the transformation does not introduce an anomaly, in a quantum theory the creation and annihilation operators of relativistic and non-relativistic fields can be expressed in terms of each other through a Bogoliubov transformation.
In Minkowski space-time this transformation consists only of a multiplication by a time-dependent phase factor and thus leads to the same vacuum for relativistic and non-relativistic particles.
While the representation as a Bogoliubov transformation still works in FLRW space-time, the freedom of choice of the mode function does not allow for a similarly general statement about the vacuum state.
		
The effective low energy theory for the non-relativistic fields does at leading orders not account for any particle number changing processes (e.g. a 4 to 2 scattering in a $\phi^4$ theory).
Such processes would introduce further imaginary terms into the effective potential to account for the loss of slow non-relativistic particles to higher velocities.
One way these terms can be included is by matching $T$-matrix elements of relativistic and non-relativistic theory in the limit of vanishing ingoing 3-momenta as done by Braaten, Mohapatra, and Zhang \cite{BraatenLossTerms}.

In the present paper we have considered the transformation from relativistic to non-relativistic fields as a rewriting of the microscopic action $S[\phi] \to S[\psi]$. However, because it is a linear transformation, one may equally well relate the one-particle irreducible or quantum effective actions $\Gamma[\phi] \to \Gamma[\psi]$ which are now functions of field expectation values.

What we have not addressed fully here is the question of anomalies. Because the transformation between real relativistic and complex non-relativistic fields is linear it is plausible that it remains anomaly free, but this may need a more detailed investigation in the future.

Finally, it would also be interesting to extend the discussion presented here to fermionic fields, and specifically to study the non-relativistic limit of Majorana fermions by similar means. 

\begin{acknowledgments}
L.H.H. acknowledges useful discussions with Tobias Haas.
This work is supported by the Deutsche Forschungsgemeinschaft (DFG, German Research Foundation) under Germany's Excellence Strategy EXC 2181/1 - 390900948 (the Heidelberg STRUCTURES Excellence Cluster), Collaborative Research Centre "SFB 1225 (ISOQUANT)", as well as FL 736/3-1.
\end{acknowledgments}

\appendix
\section{Canonical transformation of the Hamiltonian \label{app:CanonicalTrans}}
From the condition \eqref{eq:ansatzPhiPi} we find that the Hamiltonian has to be of the form
\begin{align}
\mathcal{H}_{\text{old}} = \frac{1}{2} \pi A \pi + \frac{1}{2} \phi B \phi.
\end{align}
The Hamiltonian of the new fields after canonical transformation is given as
\begin{align}
\mathcal{H}_{\text{new}}(\varphi_1, \varphi_2) = \mathcal{H}_{\text{old}}(\phi, \pi) + \frac{\partial F_2}{\partial t}.
\end{align}
From \eqref{eq:psiCtoOthers} and \eqref{eq:pitoOthers} we can read off the generating function of the transformation
\begin{equation}
\begin{split}
F_2(\phi, \varphi_2, t) = & -\frac{1}{2} \varphi_2 \left( \frac{\Im(\alpha\gamma)}{\Re(\alpha\gamma)} \right) \varphi_2 - \frac{1}{2} \phi \left( \frac{\Im(\alpha)}{\Re(\alpha\gamma)} \right) \phi \\
& + \frac{1}{\sqrt{2}} \varphi_2 \left( \frac{1}{\Re(\alpha\gamma)} \right) \phi.
\end{split}
\end{equation}
Substituting the new variables into the old Hamiltonian gives
\begin{align}
\begin{split}
\mathcal{H}_{\text{old}} = & \frac{1}{2}\varphi_1 [B\Re(\gamma) - 2\Im(\alpha)\Re(\alpha\dot{\gamma})] \varphi_1  \\ 
& + \frac{1}{2} \varphi_2 [B\Im(\gamma) + 2\Re(\alpha)\Im(\alpha\dot{\gamma})] \varphi_2 \\
&+ \varphi_2 [\Re(\alpha^2 \dot{\gamma})] \varphi_1,
\end{split}
\end{align}
and doing so for $\partial_t F_2$ yields
\begin{equation}
\begin{split}
\partial_t F_2 &= \varphi_2 \left[ -\frac{1}{2}  \partial_t \left( \frac{\Im(\alpha\gamma)}{\Re(\alpha\gamma)} \right)- \Im^2(\alpha\gamma) \partial_t \left( \frac{\Im(\alpha)}{\Re(\alpha\gamma)} \right) \right. \\
& \quad\quad\quad\left. + \Im(\alpha\gamma) \partial_t \left( \frac{1}{\Re(\alpha\gamma)} \right) \right] \varphi_2\\
&+ \varphi_2 \left[ -2\Im(\alpha\gamma)\Re(\alpha\gamma) \partial_t \left( \frac{\Im(\alpha)}{\Re(\alpha\gamma)} \right) \right.\\
& \quad\quad\quad \left. + \Re(\alpha\gamma) \partial_t \left( \frac{1}{\Re(\alpha\gamma)} \right) \right] \varphi_1\\
&+ \varphi_1 \left[ -\Re^2(\alpha\gamma) \partial_t \left( \frac{\Im(\alpha)}{\Re(\alpha\gamma)} \right) \right] \varphi_1.
\end{split}
\end{equation}
Adding these gives us the Hamiltonian for the new variables which assumes the form
\begin{equation}
\begin{split}
\mathcal{H}_{\text{new}} = & \frac{1}{2} \varphi_2 \left( -\partial_t \arg(\alpha) + B\Re(\gamma) \right) \varphi_2 \\
& + \frac{1}{2} \varphi_1 \left( -\partial_t \arg(\alpha) + B\Re(\gamma) \right) \varphi_1,
\end{split}
\end{equation}
or written in terms of $\psi$ and $\psi^*$,
\begin{align}
\mathcal{H} = -\psi^* \left( \partial_t \arg(\alpha) - B\Re(\gamma) \right) \psi.
\end{align}
This agrees nicely with equation \eqref{eq:EOMpsi}.

\section{Analytical solution for $\gamma$ in FLRW space-time \label{app:gammaExact}}
In FLRW space-time equation \eqref{eq:gammaDEQNonFlat} takes the form
\begin{equation}
\partial_t \gamma + i m^2\mathcal{P}_a^2 \gamma^2 - 3H \gamma - i = 0~.
\end{equation}
We now assume there is an solution for gamma which is analytical in $H$ and hence can be expressed as a power series
\begin{equation}
\gamma = \sum_{n=0}^{\infty} f_n(\mathcal{P}_a) H^n~.
\end{equation}
We further assume that the ratio
\begin{equation}
\Xi = - \frac{\dot{H}}{H^2} \label{eq:XiDefinition}
\end{equation}
(directly related to the deceleration parameter, $q=\Xi-1$) is constant. 
Now we can look at the individual terms of the differential equation at each order in $H$.
The constant term only contributes at order $H^0$.
The linear term shifts the coefficient functions $f_n$ by one order,
\begin{equation}
-3H\gamma = -3\sum_{n=1}^{\infty} f_{n-1}(\mathcal{P}_a) H^n.
\end{equation}
The quadratic term contains all coefficient functions up to the order we are looking at,
\begin{equation}
\begin{split}
& i m^2 \mathcal{P}_a^2 \gamma^2 = i m^2 \mathcal{P}_a^2 \sum_{n=0}^{\infty} \sum_{k=0}^{n} f_{k}(\mathcal{P}_a) f_{n-k}(\mathcal{P}_a) H^n,\\
&= i m^2 \mathcal{P}_a^2 \sum_{n=0}^{\infty}  H^n \\
& \times \left[ (2-\delta_{n0}) f_0(\mathcal{P}_a) f_n(\mathcal{P}_a) + \sum_{k=1}^{n-1} f_{k}(\mathcal{P}_a) f_{n-k}(\mathcal{P}_a) \right].
\end{split}
\end{equation}
To calculate the time derivative of the $f_n(\mathcal{P}_a)$ we use
\begin{equation}
\partial_t \mathcal{P}_a = H (\mathcal{P}_a^{-1} - \mathcal{P}_a),
\end{equation}
which we can then write as
\begin{equation}
\partial_t f_n(\mathcal{P}_a) = H (\mathcal{P}_a^{-1} - \mathcal{P}_a) f^\prime_n (\mathcal{P}_a),
\end{equation}
where $f^\prime_n$ is to be understood as the derivative $\partial_x f_n(x)$ evaluated at the given argument.
With assumption \eqref{eq:XiDefinition} we get
\begin{equation}
 \partial_t H^n = - n \Xi H^{n+1}.
\end{equation}
This means that a time derivative always increases the order in $H$ by one.
Thus the time derivative term in the differential equation can be written as
\begin{equation}
\partial_t \gamma = \sum_{n=1}^{\infty} H^n [(\mathcal{P}_a^{-1} - \mathcal{P}_a) f^\prime_{n-1}(\mathcal{P}_a) - (n-1) \Xi f_{n-1}(\mathcal{P}_a)].
\end{equation}
At zeroth order in $H$ only the constant and quadratic terms contribute and we get
\begin{equation}
i m^2 \mathcal{P}_a^2 f_0(\mathcal{P}_a)^2 - i = 0,
\end{equation}
which is solved by
\begin{equation}
f_0(\mathcal{P}_a) = m^{-1} \mathcal{P}_a^{-1}.
\end{equation}
The negative solution does not lead to a valid transformation as it makes it impossible for $\alpha$ to fulfil condition \eqref{eq:alphaCondition}. At order $H^n$, $n\geq 1$, we have a recursive solution for the $f_n$,
\begin{equation}
\begin{split}
f_n = & \frac{i}{2 m \mathcal{P}_a} {\bigg [} (\mathcal{P}_a^{-1} - \mathcal{P}_a) f^\prime_{n-1} - (3 + (n-1) \Xi) f_{n-1} \\
&  + i m^2 \mathcal{P}_a^2 \sum_{k=1}^{n-1} f_k f_{n-k} {\bigg ]}.
\end{split}
\end{equation}
By induction one can prove using the above formula that for $n\in \mathbb{N}$
\begin{equation}
\Im(f_{2n}(\mathcal{P}_a)) = 0, \quad\quad\quad \Re(f_{2n+1}(\mathcal{P}_a)) = 0.
\end{equation}
As the Lagrangian for the free non-relativistic field only depends on $\Re(\gamma)$, this implies that it is an even function of the Hubble rate $H$.

\vspace{10mm}

\bibliography{sources.bib}
	
\end{document}